\documentclass[pra,twocolumn,preprintnumbers,amsmath,amssymb,superscriptaddress,showpacs]{revtex4}

\usepackage{graphicx}
\usepackage{dcolumn}
\usepackage{bm}

\begin{document}


\title{Full Bloch sphere teleportation of spinor Bose-Einstein condensates and spin ensembles}
\author{Alexey N. Pyrkov}
\affiliation{Institute of Problems of Chemical Physics RAS, Acad. Semenov av., 1, Chernogolovka, 142432, Russia}

\author{Tim Byrnes}
\affiliation{National Institute of Informatics, 2-1-2
Hitotsubashi, Chiyoda-ku, Tokyo 101-8430, Japan}

\date{\today}

\begin{abstract}
Quantum teleportation is the transfer of quantum information between two locations 
by the use of shared entanglement.  Current teleportation schemes broadly fall under one of two categories, of either qubit or continuous variables teleportation. Spin coherent states on spin ensembles can be teleported under the continuous variables approximation for states with small deviations from a given polarization.  However, for spin coherent states with large deviations, no convenient teleportation protocol exists. Recently, we introduced a teleportation scheme where a spin coherent state lying on the equator of the Bloch sphere is teleported between distant parties (Pyrkov \& Byrnes New. J. Phys. 16, 073038 (2014)).  Here we generalize the protocol to a spin coherent state with an arbitrary position on Bloch sphere.  Our proposed scheme breaks classical bounds based on communication in the unconditional case and quantum state estimation with postselection. 
\end{abstract}


\maketitle


\section{Introduction}

Quantum teleportation \cite{bennett93,vaidman94} is one of the most enigmatic protocols introduced so far in the field of quantum information, allowing for the transfer of quantum information assisted by an entangled state.  Teleportation has been experimentally realized using a variety of systems ranging from those using photons \cite{bouwmeester97,boschi98,furusawa98}, atoms \cite{riebe04,barrett04,olmschenk09}, and hybrid systems \cite{sherson06,chen08}. For larger objects involving many particles, teleportation is typically more difficult due to the quantum-to-classical transition.  This is attributed to the fast decoherence such macroscopic objects suffer, where the large number of particles accelerate the rate of decoherence \cite{schlosshauer07,zurek03}. 
Nevertheless, there have been several experimental demonstrations of entanglement generation and teleportation  between macroscopic objects \cite{fickler12,sherson06,julsgaard01,chou05,krauter12,bao12}. For example, spin polarized atomic ensembles were used to form continuous variables (CV) \cite{braunstein05} and entanglement was generated between two ensembles \cite{julsgaard01}.  More recently, teleportation was accomplished both in the CV framework \cite{krauter12} and using spin wave states \cite{bao12} between two atomic ensembles.

What allows for these experiments involving macroscopic objects to succeed, despite the large number of particles involved?
To answer this question, it is important to consider the types of quantum states that are 
involved.  While these experiments are macroscopic in the sense that there are a large number of particles involved, in terms of the quantum mechanical degrees of freedom, typically a relatively small fraction of the total Hilbert space is used. For example, for CV teleportation using ensembles, typically the spins are polarized in the $ S^x $-direction, and the remaining spin directions form effective position and momentum variables \cite{braunstein05}
\begin{align}
x & = \frac{S^y}{\sqrt{N}} \nonumber \\
p & = \frac{S^z}{\sqrt{N}} ,
\label{xandp}
\end{align}
where $ N $ is the number of atoms in the ensemble, and $ S^{x,y,z} $ is the total spin of the ensemble in the $ x,y,z $ directions. The teleported states are then small displacements from the completely $ S^x $-polarized state on the Bloch sphere 
\cite{krauter12}. For large displacements the Holstein-Primakoff approximation breaks down, 
and (\ref{xandp}) cannot be regarded as quadrature variables. In the experiment of Ref. \cite{krauter12}, the mean displacement from the vacuum state is typically less than 10 spin flips from the completely $ S^x $-polarized state, a tiny fraction of the $ 2^N $-dimensional Hilbert space used with $ N \sim 10^{12} $ atoms involved. For the protocol in Ref. \cite{bao12}, the teleported state is a single particle excitation from the completely spin polarized state
\begin{align}
\frac{1}{\sqrt{N}} \sum_n e^{i \bm{k} \cdot \bm{r}_n} \sigma^+_n | \downarrow \downarrow  \dots \downarrow  \rangle 
\end{align}
where $ \bm{k} $ is the momentum of the spin wave state encoding the state to be teleported, $ \bm{r}_n $ is the physical location of the $ n $th spin, and $ \sigma^+_n $ flips the $n$th spin on the ensemble. Here only a single spin flip from the completely polarized state is involved, in this case with $ N \sim 10^8 $.  

Both the spin wave state and the coherent state are known to be relatively robust in the presence of decoherence \cite{lukin01}.  This is unlike for example a Schrodinger cat state 
\begin{align}
\alpha | \uparrow \uparrow  \dots \uparrow  \rangle + \beta | \downarrow \downarrow  \dots \downarrow  \rangle  ,
\label{catstate}
\end{align}
where $ \alpha, \beta $ are arbitrary coefficients to be teleported with $|\alpha|^2 + |\beta^2| =1 $.  The state (\ref{catstate}) has a decoherence rate that is enhanced from the single particle value by a factor $ N^2 $, where $ N $ is the number of particles in the ensemble. While it is not difficult to propose a teleportation based on states such as (\ref{catstate}), this would never succeed under realistic circumstates due to this enhanced decoherence.  The success of the CV and spin wave approaches experimentally can therefore be attributed are relatively robust nature of the states involved.   

The small portion of the Hilbert space involved in the approaches of Refs. \cite{krauter12,bao12} may suggest that teleporting a macroscopic state involving all $ 2^N $ states in Hilbert space is a difficult task.  While there have been works showing that in principle a general state could be teleported by generalizing qubit teleportation techniques \cite{chen06}, for macroscopic objects these are prohibitively complicated as it requires microscopic control of each particle in the object.  In order to be experimentally realizable, a prospective macroscopic teleportation protocol should involve a relatively small number of steps and be robust under decoherence.  In this paper, we propose precisely such a protocol.  

Recently we introduced a protocol for teleporting a spin coherent state between two spinor Bose-Einstein condensates (BECs) by the use of shared entanglement \cite{pyrkov14}. The type of quantum state that is teleported takes the form
\begin{equation}
\label{becqubit}
|\alpha,\beta\rangle\rangle\equiv\frac{1}{\sqrt{N!}}(\alpha a^\dagger+\beta b^\dagger)^{N}|0\rangle,
\end{equation}
where creation operators for the two hyperfine states of the BEC $ a^\dagger, b^\dagger $ obey bosonic commutation relations, and $ N $ is the number of bosons in the BEC. Due to the equivalence of the algebra of total spin operators, the scheme equally applies to spin ensembles \cite{abdelrahman14}
\begin{equation}
\label{spinensqubit}
|\alpha,\beta \rangle \rangle \equiv \bigotimes_n (\alpha |\uparrow \rangle_n  +\beta |\downarrow \rangle_n ).
\end{equation}
It can be seen that the state (\ref{becqubit}) and (\ref{spinensqubit}) involves the full Hilbert space in each case, depending upon the choice of the parameters $ \alpha, \beta $.  
Our scheme introduced in Ref. \cite{pyrkov14} was restricted to states that lie on the 
equator of the Bloch sphere.  Using the standard Bloch sphere parametrization $ \alpha = \cos \frac{\theta}{2} e^{-i \phi/2} , \beta = \sin \frac{\theta}{2} e^{i \phi/2}  $, this corresponds to the restriction $ \theta = \pi/2 $.  In this paper we extend the scheme to an arbitrary spin coherent state on the Bloch sphere.  This is done by generalizing the previous protocol by using two different entangling channels for the $ \theta $ and $ \phi $ spherical coordinates. The number of operations we use in the protocol is independent of the number of atoms $ N $ in the BEC or ensemble.  Despite the increased number of spin ensembles used, remarkably only a binary classical correction variable needs to be sent.  We show that in terms of the amount of information that is transmitted, it exceeds the classical values, which may be attributed to the entanglement between Alice and Bob.

It is clear from the structure of (\ref{becqubit}) and (\ref{spinensqubit}) that the quantum information is duplicated $ N $ times in a spin coherent state. Such a structure is attractive from a quantum information standpoint since it adds a robustness via duplication of the information.  Thus unlike single qubit systems (not 
involving quantum error-correction) where a single quanta of external noise can destroy the quantum 
information, even a loss of some fraction of the number of particles in (\ref{becqubit}) does not
result in complete loss of the quantum information, it merely contributes to a diminished signal amplitude. Previously we proposed a method of using such spin coherent states 
for quantum computational purposes, exploiting the analogous properties to standard qubits 
\cite{byrnes12}.  While the approach has similarities with CV approaches, by using all states on the Bloch sphere - not just those in the vicinity of a given polarization direction (e.g. $ S^x $) - one can no longer treat the system in terms of bosonic modes, and thus goes beyond the CV framework.  The present proposal adds another protocol to the list of quantum information tasks that are possible with the approach \cite{byrnes12,byrnes14,pyrkov14,byrnes12b,pyrkov12,gross12}.  


\section{The teleportation protocol}

\label{secfull}

The aim of the teleportation protocol will be to transfer an unknown spin coherent state $ | \alpha, \beta \rangle \rangle $ between two parties by the use of shared entanglement. We continue the tradition and call the heroes of our protocol as Alice (sender) and Bob (receiver). We will allow Alice to have ancilla BECs or ensembles, initially in a spin coherent state, to assist in the teleportation protocol.  We parameterize the spin coherent state to be sent as $ \alpha = \cos \frac{\theta}{2} e^{-i \phi/2} , \beta = \sin \frac{\theta}{2} e^{i \phi/2}  $, with $ \theta=[0,\pi], \phi = [-\pi,\pi] $. Without loss of generality, we shall restrict our presentation to the BEC case (\ref{becqubit}), as the translation to the spin ensemble case can be performed straightforwardly \cite{abdelrahman14}.  Following previous works \cite{byrnes12} we shall call the spin coherent state (\ref{becqubit}) a ``BEC qubit'' due to the analogous properties this case with standard qubits.  We emphasize that the BEC qubit (\ref{becqubit}) is strictly not a genuine qubit as it is clearly not a two level system. However, the analogy will be useful as it contains the same information as a standard qubit in terms of the parameters $ \alpha, \beta $, but with a different encoding.

\begin{figure*}
\scalebox{0.43}{\includegraphics{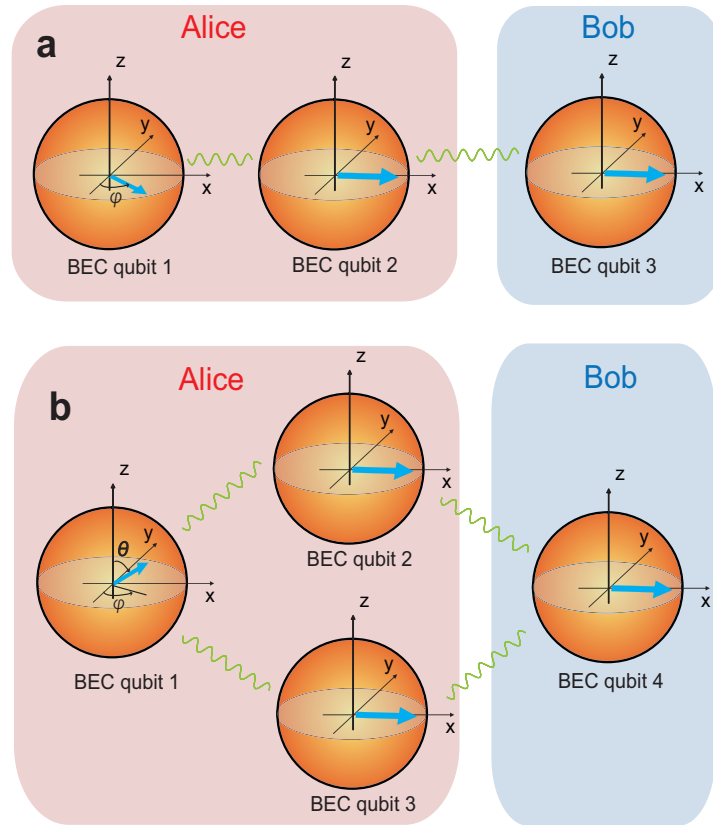}}
\caption{The schematic teleportation protocol. The aim is to send an unknown spin coherent state on BEC qubit 1 owned by Alice to Bob.  (a) The equatorial teleportation protocol as
introduced in Ref. \cite{pyrkov14}.  (b) The full Bloch sphere teleportation protocol considered in this paper. Wiggly lines denote entanglement created using an $ S^z S^z $ interaction.
\label{fig0} }
\end{figure*}

In the previously proposed equatorial state teleportation of Ref. \cite{pyrkov14}, 
one additional ancilla BEC qubit was used by Alice, as shown in  Fig. \ref{fig0}(a). An outline of the equatorial protocol is as follows.  Initially Alice has an unknown spin coherent state with angle $ \phi $ along the equator of the Bloch sphere ($ \theta =\pi/2 $).  Alice's ancilla BEC qubit and Bob's initial state is in the $S^x $ polarized state, i.e. $ \phi = 0, \theta = \pi/2 $.  Entanglement is produced between BEC qubits 1+2 and 2+3 (as defined in  Fig. \ref{fig0}(a)) with an $ S^z S^z $ interaction applied for a time $ t = 1/\sqrt{2N} $.  After application of an Hadamard gate on BEC 1, a number measurement is made on BECs 1 and 2.  A binary variable based on the measurement result is sent between Alice and Bob, from which Bob makes a classical correction to complete the teleportation procedure. 

An intuitive explanation how the equatorial teleportation scheme works is as follows. An $ S^z S^z $ interaction is known to create quantum correlations between spin coherent states lying on the equator of the Bloch sphere and $S^z $-eigenstates on the other BEC qubit \cite{byrnes13}. In the teleportation protocol the entangling step between BEC qubit 1 and 2 produces $\phi_1$-$S^z_2 $ correlations.   By producing $ S^z S^z $ entanglement between BEC qubit 2 and 3, this maps the correlations back from the $ S^z $ eigenstates on BEC qubit 2 to equatorial spin coherent states on BEC 3 ($S^z_2 $-$\phi_3$ correlations).  In total this creates $ \phi_1 $-$S^z_2 $-$\phi_3$   correlations, creating entanglement between equatorial spin coherent states on 
BEC qubit 1 and BEC qubit 3.  By measuring out BEC qubits 1 and 2, the original state on BEC qubit 1 is teleported to BEC qubit 3.  Due to the randomness of the measurement process, Alice needs to send her measurement result to Bob so that he may make the necessary corrections.  

Our full Bloch sphere teleportation protocol generalizes this procedure by using two ancilla BEC qubits, labeled by BEC qubits 2 and 3 (Fig. \ref{fig0}(b)).  The basic idea is to use two channels for the two variables $ \theta, \phi $.  BEC qubit 2 contains the correlations for the $ \phi $ degree of freedom, such that after the entangling steps are performed, $ \phi_1 $-$S^z_2 $-$\phi_4$ correlations are produced.  BEC qubit 3 takes care of the $ \theta $ degree of freedom, mediating $ \theta_1 $-$S^z_3 $-$\theta_1$ correlations.  As before, once the suitable entanglement is created, BEC qubits 1, 2, and 3 are measured out in the number basis. A binary variable is based on Alice's measurement result is sent, from which Bob performs a classical correction procedure.  
  
We now describe  the protocol in detail. Before the teleportation, the initial state of whole system is
\begin{align}
\label{initial2}
& |\cos \frac{\theta}{2} e^{-i \phi/2} ,\sin \frac{\theta}{2} e^{i \phi/2} \rangle\rangle_1  |\frac{1}{\sqrt2},\frac{1}{\sqrt2}\rangle\rangle_2
\nonumber \\
& \times |\frac{1}{\sqrt2},\frac{1}{\sqrt2}\rangle\rangle_3  |\frac{1}{\sqrt2},\frac{1}{\sqrt2}\rangle\rangle_4 .
\end{align}
The aim is to teleport an unknown state on BEC qubit 1 (Alice) to BEC qubit 4 (Bob).  The protocol then follows the following sequence.
\begin{enumerate}
\item Apply the entangling gate $ -S^z_1 S^z_2$ for a time $ \tau_2 =1/\sqrt{2N} $ between BEC qubits 1 and 2
\item Apply a Hadamard gate on BEC qubit 1 of the form $a^\dagger\rightarrow \frac{1}{\sqrt{2}}(a^\dagger+b^\dagger)$, $b^\dagger\rightarrow \frac{i}{\sqrt{2}}(a^\dagger-b^\dagger)$
\item Apply the entangling gate $ -S^z_1 S^z_3 $ on BEC qubits 1 and 3 for a time $\tau_3=1/\sqrt{8N} $  
\item Apply a Hadamard gate on BEC qubit 1 of the form $a^\dagger\rightarrow \frac{1}{\sqrt{2}}(a^\dagger+i b^\dagger)$, $b^\dagger\rightarrow \frac{i}{\sqrt{2}}(a^\dagger-ib^\dagger)$
\item Apply the entangling gate $ S^z_2 S^z_4$ on BEC qubits 2 and 4 for a time $T_2=1/\sqrt{2N} $
\item Apply a Hadamard gate on BEC qubit 4 in the form $a^\dagger\rightarrow \frac{1}{\sqrt{2}}(a^\dagger+ib^\dagger)$, $b^\dagger\rightarrow \frac{1}{\sqrt{2}}(a^\dagger-ib^\dagger)$ 
\item Apply the entangling gate $ S^z_3 S^z_4$ on BEC qubits 3 and 4 for a time $T_3=1/\sqrt{8N} $
\item Measure BEC qubits 1, 2 and 3 in the $ | k \rangle $ basis
\item Classically transmit the binary measurement result $ \sigma_1=\mbox{Sgn}(2 k_1-N) $ to Bob and perform the transformations 
$ \phi \rightarrow \phi + \pi(1- \sigma_1)/2$ and $ \theta \rightarrow \sigma_1 \theta + \pi/2 $.
\end{enumerate}

\begin{figure}
\scalebox{0.3}{\includegraphics{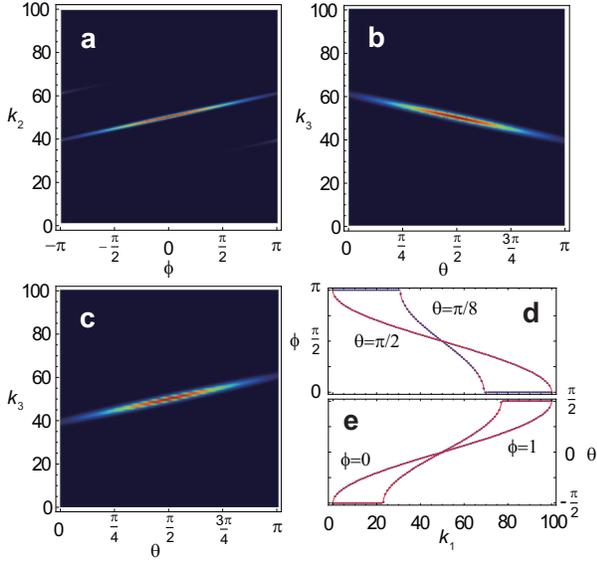}}
\caption{
Full Bloch sphere teleportation probability distributions of measurements on Alice's qubits $ p(k_1,k_2,k_3) $ for (a) $ k_1 = N,$ $ k_3 = N/2$, $ \theta= \pi/2 $; (b) $ k_1 = N$, $ k_2 = N/2$, $ \phi= 0 $; (c) $ k_1 = 0,$ $ k_2 = N/2$, $ \phi= \pi $. Position of probability density peaks for $ k_2 = N/2 $, $ k_3=N/2 $ for constant (d) $\theta $ and (e) $ \phi $. Red lines show analytical formula and dots are numerically obtained results. All calculations presented with no dephasing ($ \gamma = 0 $) and $N=100$. 
\label{fig1S} }
\end{figure}

\begin{figure}
\scalebox{0.3}{\includegraphics{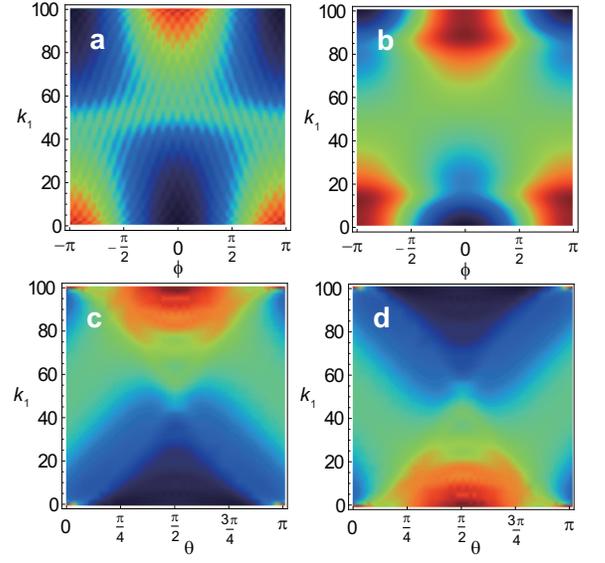}}
\caption{The probability of obtaining a particular outcome $ k_1 $ in the teleportation protocol defined as $ \sum_{k_2,k_3} p(k_1,k_2,k_3) $ for (a) $ \theta = \pi/2 $ (b) $ \theta = \pi/4 $ (c) $ \phi=0 $ (d) $ \phi=\pi $ as a function of the initial states as shown.  
All calculations are with $N=100$. 
\label{fig1S2} }
\end{figure}

The state of the BECs then evolve as follows. After Step 1 the state is
\begin{align}
& \frac{1}{\sqrt{2^N}} \sum_{k_2=0}^N \sqrt{C_N^{k_2}} \nonumber \\
& \times |\cos \frac{\theta}{2} e^{-i (\phi/2 -s_2 \tau_2)}, \sin \frac{\theta}{2} e^{i (\phi/2 -s_2 \tau_2)} \rangle\rangle_1 |k_2 \rangle 
\end{align}
where $s_j=2 k_j-N$. Following Step 2 we have 
\begin{align}
\frac{1}{2^N} \sum_{k_2=0}^N \sqrt{C_N^{k_2}} |\cos(\phi/2-s_2 \tau_2) e^{i\frac{\theta}{2}}-i\sin(\phi/2-s_2 \tau_2) e^{-i\frac{\theta}{2}}  ,\nonumber \\ 
\cos(\phi/2-s_2 \tau_2) e^{-i\frac{\theta}{2}}-i\sin(\phi/2-s_2 \tau_2) e^{i\frac{\theta}{2}} \rangle\rangle_1 |k_2 \rangle. 
\end{align}
By performing Step 3 we connect the amplitude of the unknown state with the measurement result of the third BEC qubit using the entangling gate
\begin{align}
&\frac{1}{\sqrt{8^{N}}}  \sum_{k_2,k_3=0}^N \sqrt{C_N^{k_2}C_N^{k_3}} \nonumber \\
& \times |\cos(\frac{\phi}{2}-s_2 \tau_2) e^{i(\frac{\theta}{2}+s_3 \tau_3)}-i\sin(\frac{\phi}{2}-s_2 \tau_2) e^{-i(\frac{\theta}{2}-s_3 \tau_3)},  \nonumber \\
& \cos(\frac{\phi}{2}-s_2 \tau_2) e^{-i(\frac{\theta}{2}+s_3 \tau_3)}-i\sin(\frac{\phi}{2}-s_2 \tau_2) e^{i(\frac{\theta}{2}-s_3 \tau_3)} \rangle\rangle_1 \nonumber \\
& \times |k_2 \rangle|k_3 \rangle. 
\label{step4}
\end{align}
After Step 4 we have the state
\begin{align}
 \frac{1}{2^N}& \sum_{k_2,k_3=0}^N \sqrt{C_N^{k_2}C_N^{k_3}} | A_{k_2 k_3} (\theta,\phi), B_{k_2 k_3}  (\theta,\phi) \rangle \rangle_1 | k_2\rangle | k_3 \rangle ,
\end{align}
where 
\begin{align}
A_{k_2 k_3} (\theta,\phi) = & \cos (\phi/2 - s_2 \tau_2) \sin ( \pi/4 + \theta/2 + s_3 \tau_3)  \nonumber \\
& + i \sin (\phi/2 - s_2 \tau_2) \sin ( \pi/4 - \theta/2 + s_3 \tau_3) , \nonumber \\
B_{k_2 k_3}  (\theta,\phi) = & -\cos (\phi/2 - s_2 \tau_2) \sin ( \pi/4 - \theta/2 - s_3 \tau_3) \nonumber \\
& -i \sin (\phi/2 - s_2 \tau_2) \sin ( \pi/4 + \theta/2 - s_3 \tau_3) .
\end{align}
At this point correlations exist between the $ S^z $ eigenvalues of the second and third BEC qubits and the spherical coordinates of the unknown state. The purpose of Steps 4 to 7 is to correlate the state of BEC qubit 4 to the measurement outcomes $ k_2 $ and $ k_3 $. The final state before the measurement step (after step 7) is 
\begin{align}
\frac{1}{2^N} \sum_{k_2,k_3=0}^N & \sqrt{C_N^{k_2}C_N^{k_3}} | A_{k_2 k_3} (\theta,\phi), B_{k_2 k_3} (\theta,\phi) \rangle \rangle_1 | k_2\rangle | k_3 \rangle \nonumber \\
& \times |\cos(s_3 T_3) e^{-i s_2 T_2}  , \sin(s_3 T_3) e^{i s_2 T_2} \rangle\rangle_4. 
\label{finalstates}
\end{align}
The probability distribution of the measurement is
\begin{align}
p(k_1,k_2,k_3)= & \frac{1}{4^N}C_N^{k_1}C_N^{k_2}C_N^{k_3} \nonumber \\
& \times | A_{k_2 k_3}  (\theta,\phi) |^{2 k_1}  | B_{k_2 k_3} (\theta,\phi) |^{2N-2k_1}.
\label{finalprob}
\end{align}

If we consider separately the connection between $k_2$ and $\phi$ and connection between $k_3$ and $\theta$ the probability distribution gives results which are similar to the equatorial teleportation case~\cite{pyrkov14}. In Fig. \ref{fig1S}(a) we see that there is a sharp correlation between  $ \phi $ in the initial state and the measurement outcomes $ k_2 $, with little dependence on $ \theta $.  Figures \ref{fig1S}(b)(c) show the sharp correlations between $ \theta $ and $ k_3 $. The general features of Figs. \ref{fig1S}(b) are qualitatively unchanged for $ | \phi | < \pi/2 $, with this swapping over to Fig. \ref{fig1S}(c)  for $ | \phi | > \pi/2 $ ($ \mbox{mod} 2 \pi $).  Thus there is a swap in the sign of the gradient according to which hemisphere Alice's state lies in.  As $ k_1 $ is varied, the peak positions shift as in the equatorial case.  Figures \ref{fig1S}(d)(e) show the position of the peaks with varying $ k_1 $.  The numerically determined points have perfect agreement with the expressions
\begin{align}
\phi_{\mbox{\tiny peak}} & = \arccos \left( \frac{s_1}{N\sin \theta} \right) \nonumber \\
\theta_{\mbox{\tiny peak}} & = \arcsin \left( \frac{s_1}{N\cos \phi} \right) .
\end{align}
We may thus deduce that with very high probability the measured result will occur at
\begin{align}
s_2 \tau_2 & = \frac{\phi}{2} - \frac{1}{2} \arccos \left( \frac{s_1}{N \sin \theta} \right)\label{s2tau} \\
s_3 \tau_3 & = -\mbox{Sgn}(\cos \phi) \frac{\theta}{2}  - \frac{1}{2} \arcsin \left( \frac{s_1}{N \cos \phi} \right).
\label{linkage}
\end{align}
Thus Bob's qubit (\ref{finalstates}) becomes correlated with the original state on Alice's BEC qubit.  

It may appear that the $ \theta $ and $ \phi $ dependence of the correction terms (the second terms on the right hand side of (\ref{s2tau}) and (\ref{linkage}) and the $ \mbox{Sgn}(\cos \phi) $ coefficient in (\ref{linkage})) would mean that it is necessary to know what Alice's initial state is in order to perform the classical correction. Fortunately, to a good approximation this is not the case.  Figure \ref{fig1S2} shows the total probability distribution of the $ k_1 $ measurement outcome.  It shows that it is generally always centered around either $ k_1 = 0,N $. On the equator (Fig. \ref{fig1S2}(a)), the distribution resembles the equatorial teleportation case \cite{pyrkov14}.  Going off the equator (Fig. \ref{fig1S2}(b)), the distribution slightly broadens towards the center, but the correction term as shown in Fig. \ref{fig1S}(d) becomes more likely to be at its extreme values. Figs. \ref{fig1S2}(c)(d) show the probabilities of the $ k_1 $ outcomes as a function of $ \theta $. This shows that there is essentially no $ \theta $ dependence, but correlations exists between which hemisphere $ \phi $ lies in and the value of $ k_1 $, such that we may say that $ \mbox{Sgn} (\cos \phi) = \mbox{Sgn} (s_1) $. These show that in practice the correction factor is highly likely to be at the extremal values of the $ \arccos $ and $ \arcsin $ functions, and may be approximated by $ \pi(1- \mbox{Sgn}(s_1))/2 $ and $ \pi \mbox{Sgn}(s_1)/2 $ instead. 
Taking into account that $ \sin \theta \ge 0 $ for $ 0  \le \theta \le \pi $, an approximate relation may then be obtained
\begin{align}
s_2 \tau_2 & = \frac{\phi}{2} - \frac{\pi(1- \mbox{Sgn}(s_1))}{4} \nonumber \\
s_3 \tau_3 & = - \mbox{Sgn}(s_1) \frac{\theta}{2}  - \frac{\pi}{4}.
\label{linkage2}
\end{align}
This gives the correction rule in Step 9. This completes the full Bloch sphere teleportation and Bob has Alice's original spin coherent state to high probability.

\section{Performance and comparison to classical strategies}

\begin{figure}
\scalebox{0.35}{\includegraphics{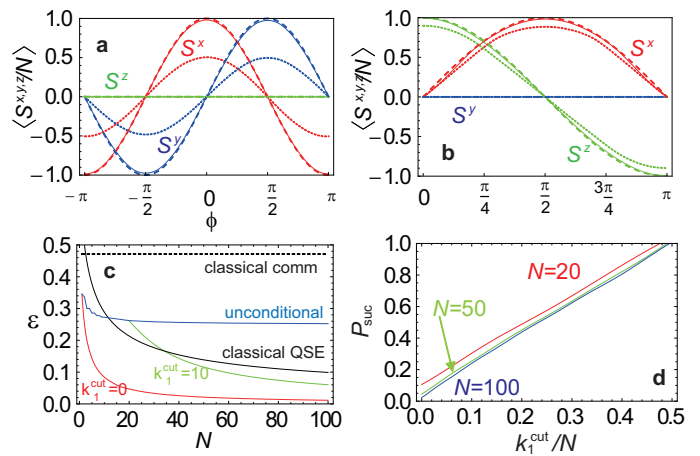}}
\caption{\label{fig2}
Performance of the teleportation scheme. (a) Average spins of Bob's BEC qubit for Alice's initial state along $ \theta = \pi/2 $ for $ k_1^{\mbox{\tiny cut}} =0 $  (solid lines) and $ N = 100 $.  Other lines corresponds to no $ k_1^{\mbox{\tiny cut}}$ parameter imposed (dotted lines), and Alice's original state (dashed lines).  (b) Average spins of Bob's BEC qubit for Alice's initial state along $ \phi = 0 $ for $ k_1^{\mbox{\tiny cut}} =0 $  (solid lines) and $ N = 100 $.  Other lines corresponds to imposing the postselection parameter is $ k_1^{\mbox{\tiny cut}} =10 $  (dotted lines), and Alice's original state (dashed lines).  (c) Error, as measured by the trace distance, of the teleportation scheme as a function of the boson number $ N $. The error bound $ \varepsilon_{\mbox{\tiny comm}} $ corresponding to 
the minimum average error due to exchange of a classical binary variable, and the error bound due to quantum state estimation (QSE) $ \varepsilon_{\mbox{\tiny QSE}} $ are shown.  (d) The success probability $P_{\mbox{\tiny suc}} $ for the postselection parameter $ k_1^{\mbox{\tiny cut}}$ with $ \theta =\pi/2 $ and $ \phi = 0 $ and the particle numbers as shown.  }
\end{figure}

We now discuss the performance of the full Bloch sphere teleportation scheme.  Figs. \ref{fig2}(a)(b) shows the average spin for the two schemes using various input states on Alice.  We see that in all cases the teleportation protocol gives Bob's BEC qubit close to the ideal behavior 
\begin{align}
\langle S^x/N \rangle & = \cos \phi \sin \theta \nonumber \\
\langle S^y/N \rangle & = \sin \phi \sin \theta \nonumber \\
\langle S^z/N \rangle & = \cos \theta .
\end{align}
There is a dependence to the performance of the protocol depending on the particular $k_1 $ outcome, due to the nature of the approximations that were used in deriving the classical correction factors (\ref{linkage2}).  Optimal results are obtained when $ k_1 $ is close to either $ 0$ or $ N $, when the correction factors (\ref{linkage2}) are identical to (\ref{linkage}).  We thus introduce a cutoff on $ k_1 $ such that only the results $ k_1 \le k_1^{\mbox{\tiny cut}} $ and $ k_1 \ge N-k_1^{\mbox{\tiny cut}} $ are kept. In an experiment this would correspond to carrying out postselection with a limited success probability of the protocol.  Figs. \ref{fig2}(b)(c) shows the dependence on the $ k_1^{\mbox{\tiny cut}} $ parameter, which shows that for $ k_1^{\mbox{\tiny cut}} = 0 $ virtually ideal teleportation is achieved.  Even without imposing the cutoff as in Fig. \ref{fig2}(a), the protocol works to a reasonable degree, although by imposing a cutoff of $ k_1^{\mbox{\tiny cut}} =0.1 N $ gives in practice very close results to the ideal case (Fig. \ref{fig2}(b)). 

An important issue when discussing quantum teleportation is evaluating the ``quantumness''
of the protocol \cite{braunstein05,braunstein00,braunstein01}. Certainly in the sense that entanglement is used to transmit information would suggest that quantum mechanics is involved.  What is however often useful, as has been the case with qubit and CV teleportation, is to compare the performance of the protocol to the best possible classical strategy of 
transmitting information. If the protocol outperforms the best possible classical strategy, then quantum mechanics must be involved. 
Assuming that Alice has no knowledge of her own state, the optimal classical strategy would be that she performes the best possible estimate of her own state, then classically transmits this to Bob.  The problem of optimally performing quantum state estimation (QSE) for $ N $ copies of a qubit has been analyzed by Massar and Popescu, where the average fidelity of the estimated state is $ (N+1)/(N+2) $ \cite{massar95}.  Since our aim is to transfer a spin coherent state, which is nothing but $ N $ copies of a qubit, we may directly use this result. It is in our case more convenient to compare trace distances rather than fidelities, thus converting the result of Ref. \cite{massar95} we obtain the classical bound
\begin{align}
\varepsilon_{\mbox{\tiny QSE}} \ge \frac{1}{\sqrt{N+2}} .
\label{popescu}
\end{align}
where our definition of the trace distance comparing Bob's state with Alice's original state is
\begin{align}
& {\cal E}(\theta, \phi) = \frac{1}{2} \Big[ (\cos \phi \sin \theta - \langle S^x_{\mbox{\tiny Bob}}/N \rangle)^2 \nonumber \\
& +(\sin \phi \sin \theta - \langle S^y_{\mbox{\tiny Bob}}/N \rangle)^2  + (\cos \theta - \langle S^z_{\mbox{\tiny Bob}}/N \rangle)^2 \Big]^{1/2} . \label{tracedistdef}
\end{align}
The expectation value in (\ref{tracedistdef}) is defined with respect to the state of Bob's qubit after the teleportation procedure. 
We shall call (\ref{popescu}) the ``QSE bound'' as it is limited by how well one can estimate an unknown state, but does not consider communication between Alice and Bob. 

In our protocol, there is another relevant bound to compare against, since Alice sends only one classical bit of information to Bob.  
Even if Alice has perfect knowledge of her own state, we may consider what the best possible classical strategy is by sending only one classical bit of information. If we assume that this is the only communication that is allowed between the two parties, then there is a limit to how well Bob can reproduce Alice's state. Given this restriction, the optimal strategy Alice may perform is to tell Bob which hemisphere her state lies in \cite{pyrkov14}. The average trace distance achievable for an even distribution of Alice's initial states can then be calculated to be
\begin{align}
\varepsilon_{\mbox{\tiny comm}}& =   \frac{1}{2 \pi} \int_0^\pi d\theta  \sin \theta \int_{-\pi/2}^{\pi/2} d\phi \nonumber  \\
& \times \frac{1}{2} 
\sqrt{ (\cos \phi \sin \theta - 1 )^2 +\sin^2 \phi \sin^2 \theta  + \cos^2 \theta }
\nonumber \\
& \approx 0.47 . 
\end{align}
We shall call the above the ``communication bound'', as it emphasizes the information is transferred between Alice and Bob using 
a single bit. 

In Fig. \ref{fig2}(c) we compare the various classical and quantum strategies. We see that the teleportation protocol outperforms the classical bounds for communication $ \varepsilon_{\mbox{\tiny comm}} $ for all values of the $ k_1^{\mbox{\tiny cut}} $ parameter, including the unconditional case. This can be attributed to entanglement between the BECs, which allows for a greater amount of information to be sent beyond the binary variable exchanged between Alice and Bob, an effect similar to superdense conding \cite{bennett92}.  In comparison to the 
quantum state estimation bound $ \varepsilon_{\mbox{\tiny QSE}}  $ we find that our teleportation protocol requires postselection to beat the classical bound for QSE.  For the optimal case of $ k_1^{\mbox{\tiny cut}} =0 $, where only the extremal outcomes of the $ k_1 $ measurement are taken, the classical QSE is beaten by the quantum strategy for all $ N $.  For larger values of $ k_1^{\mbox{\tiny cut}} $, the classical correction becomes imperfect, and results in a greater error of the final teleported state. For small $ N $ the $ k_1^{\mbox{\tiny cut}} = 10 $ curve exceeds the QSE bound as it is equal to the unconditional case until $ N = 20 $.  Beyond this, it gradually performs better exceeding the classical QSE bound 
as $ N $ increases.  

The price of the higher fidelity of the teleportation is a lower success probability, as shown in Fig. \ref{fig2}(d).  Here we plot the success probability, defined as
\begin{align}
P_{\mbox{\tiny suc}} = \sum_{k_2=0}^N  \sum_{k_3=0}^N 
\sum_{{k_1\le k_1^{\mbox{\tiny cut}} \atop k_1\ge N-k_1^{\mbox{\tiny cut}}}} p(k_1,k_2,k_3) .
\end{align}
As seen in Fig. \ref{fig1S2}, most of the probability tends to be concentrated towards either $ k_1 = 0 $ or $ k_1 = N $, which is favorable for a successful teleportation. The probabilities plotted in Fig. \ref{fig2}(d) show very little variation with particle number $ N $ and Alice's initial state $ \theta, \phi $.  We see a qualitatively similar behavior to that seen with equatorial teleportation \cite{pyrkov14}, where there is a steady increase towards 1 achieved at $ k_1 = N/2 $ which is equivalent to the unconditional case. 

The superior performance of the quantum protocol can be understood qualitatively in the following way.  In the classical strategy using QSE, the location of the initial state on the Bloch sphere can be estimated with standard deviation $ S^j/N \sim 1/\sqrt{N} $. Meanwhile, the sharp correlations of (\ref{finalprob}) give variations of order $ S^j/N \sim 1/N $, which allows for smaller fluctuations of the teleported state, giving rise to a lower error $ \varepsilon $. As quantified by the classical communication bound 
$ \varepsilon_{\mbox{\tiny comm}} $, the classically transmitted variable $ \sigma_1 $ cannot directly be used to recover the precise information about Alice's state, only at best which
hemisphere the state lies in. If we compare to a classical protocol that is limited by how much information that can be transferred between Alice and Bob then our protocol easily beats the classical limits as entanglement is used as a resource.  However, beating the QSE bound is more difficult as having $ N $ copies of a state allows for increasingly better estimates of the state with $N $.  This makes beating this bound a more difficult task, requiring postselection to exceed the classical bound. 

Interestingly, the performance of the full Bloch sphere protocol as discussed in this paper appears to perform 
better in comparison to the previous equatorial teleportation \cite{pyrkov14}.  Comparison of 
Fig. \ref{fig2}(c) in this paper with Fig. 2b of \cite{pyrkov14} shows equivalent calculations for the full Bloch sphere scheme and the equatorial scheme, with the same parameters.  For the equatorial teleportation scheme with $ k_1^{\mbox{\tiny cut}} =10 $, the classical QSE bound is exceeded when $ N \approx 60 $.  On the other hand, for our current results, the classical bound is exceed at $ N \approx 35 $.  This may be explained by the fact that the QSE bounds for the two cases are different, with the equatorial QSE being lower than the full Bloch sphere QSE bound, as there is less information that requires estimation. In addition, in the full Bloch sphere case two quantum channels are used between Alice and Bob, which allows for more resources to be used. We attribute this behavior to a combination of both of these effects.

\section{Experimental Realization}

\subsection{Required Operations}

As discussed in the introduction, our protocol is suitable both for spin ensembles and spinor Bose-Einstein condensates.  In this section we will discuss the implementation with spinor Bose-Einstein condensates, as we anticipate that this will be the most suitable 
platform for performing the types of operations that are required for teleportation. Specifically, we consider an implementation using atom chips, where multiple BECs can be realized and coherently controlled. 
In our scheme several BEC spin coherent states need to be prepared, manipulated, entangled together, and measured.  Specifically, the basic
operations that our protocol assumes are the following:
\begin{itemize}
\item  Coherent spin rotations corresponding to applications of the Hamiltonians $ S^{x,y,z} $
\item  Projective measurements which collapse the coherent state onto the number basis $|k \rangle$
\item Two BEC qubit interactions corresponding to $ S^z_1 S^z_2 $
\end{itemize}
We now discuss the feasibility of each of these operations. 

Coherent spin rotations have been already experimentally realized for two-component BECs \cite{bohi09, riedel10}.  Typically the states used as the logical states of (\ref{becqubit}) are the $ | F=1,m_F=-1 \rangle $ and $ | F=2,m_F=1 \rangle $ hyperfine states \cite{treutlein06}.  Single BEC qubit control is then achieved by a combination of microwave and radio frequency pulses which in combination with the natural energy difference between these states allow for full Bloch sphere control. Another alternative is to use optical pulses via a Raman passage through an excited state \cite{abdelrahman14}.  Experimental gate times for single BEC qubit control has been realized with microwave and radio frequency pulses are in the range of $ \sim $ ms.  Using the optical approach, gate times have been estimated to be in the range of $ \sim \mu $s \cite{abdelrahman14}. 

 Projective measurements can be performed using either absorption or fluorescent imaging \cite{bucker09, andrews97b, depue00}, which destroy the BEC in the process. Other non-destructive measurement methods which leave the BEC intact such as phase contrast imaging  could also be used in the strong measurement limit \cite{ilookeke14,higbie05}. Either method is compatible with the current protocol as after the measurement the BECs are no longer used. Of the three types of operations that are required for the protocol, single BEC coherent manipulation and measurements are routinely performed experimentally.  

Two BEC qubit entanglement is yet to be demonstrated, but there are several potential methods available.  The first possibility is using mutually connected optical cavities \cite{pyrkov13}.  Such cavity based techniques are based upon the realization of strong coupling of a BEC to a cavity \cite{colombe07} which allow for the coherent exchange of a photon to the BEC. In the scheme proposed in Ref. \cite{pyrkov13}, a photon mediated $ S^z_1 S^z_2 $ interaction is produced by a fourth order excitation-relaxation process.  Another approach is using state dependent collisions generalizing the approach initially proposed for single atoms \cite{treutlein06b}. A variation of the coupled optical cavity approach but using a generalization of geometric phase gates is another possibility \cite{hussain14}.  The recent achievement of entanglement between a BEC and an atom \cite{lettner11} would suggest that a two-BEC interaction is within experimental reach. Gate times for the $ \tau, T = 1/\sqrt{2N} $ entangling times that are necessary in the teleportation protocol introduced here are estimated to be in the region of $ \sim 15 $ $ \mu$s for the cavity method \cite{rosseau14}.  This is much less than the natural dephasing times as measured in experiments, which have been measured to be in the region of seconds \cite{treutlein06,bohi09}.  However, one should take into account that the procedure to general the entanglement also creates an effective dephasing \cite{pyrkov13}.  Past estimates on the various methods suggests that even in the presence of the excess decoherence, entangled states corresponding to the $ \tau, T = 1/\sqrt{2N} $ entangling times should be possible \cite{pyrkov13,rosseau14,hussain14}.

\subsection{Decoherence}

As we are dealing with macroscopic objects such as BECs and spin ensembles, it is a natural question whether
the effects of decoherence would be amplified rendering the current protocol useless under realistic circumstances.  
While one would expect that macroscopic objects such as BECs or spin ensembles to not survive in the presence of decoherence, as discussed in the introduction, this depends critically on the type of quantum state that is involved.  This was studied in detail in Ref. \cite{byrnes13} and was shown that the $ \tau,T = 1/\sqrt{2N} $ entangled states decay were relatively robust against decoherence. Spin coherent states also have similar decoherence characteristics to standard qubits, and have no dependence on the particle number $ N $ \cite{byrnes12}.  Decoherence characteristics of the equatorial teleportation scheme were studied in \cite{pyrkov14}, and 
were found to behave favorably with $ N $.  Since our current protocol is simply an extension of the scheme, we expect the same general behavior.  

To confirm this, let us analyze the behavior of the protocol under a generic dephasing process, which can be described by the master equation
\begin{align}
\frac{d \rho(t)}{dt} = -i [H(t), \rho] - \frac{\gamma}{2} \sum_{n=1}^4 \left[ (S^z_n)^2 \rho 
-2 S^z_n \rho S^z_n + \rho  (S^z_n)^2 \right] .
\label{dephasingmaster}
\end{align}
Here $ H(t) $ refers to the Hamiltonian sequence described in Sec. \ref{secfull}.  The two BEC entangling operations $ S^z S^z $ are expected to be the main contribution to decoherence, as these take the longest time to implement, and the procedure to implement it is expected to give rise to various dephasing processes \cite{pyrkov13}.  Hence we assume that the dephasing acts during Steps 1, 3, 5, and 7 but not during single BEC qubit operations or measurement.  

To calculate the decoherence, we use a phenomenological approach to approximate the evolution of (\ref{dephasingmaster}).  Starting from a general state $ |\psi(0) \rangle = \sum_k \psi_k | k \rangle $, with $ H(t) = 0 $ in (\ref{dephasingmaster}), the state evolves to
\begin{align}
\rho(t) = \sum_{k k'} \psi_k \psi_k^* e^{-2 \gamma t (k-k')^2} | k \rangle \langle k' | .
\label{masterdyn}
\end{align}
These dynamics can be reproduced by assuming that the dephasing process produces a random phase 
\begin{align}
|\psi(\xi) \rangle = e^{i \xi S^z} | \psi \rangle = \sum_k \psi_k e^{i (2k-N) \xi} | k \rangle ,
\end{align}
where the random phase $ \xi $ is probabilistically distributed by a Gaussian with variance $ \gamma t $.  The density matrix may then be reconstructed to be 
\begin{align}
\rho(t) = \frac{1}{\sqrt{2 \pi \gamma t}} \int_{- \infty}^{\infty} d \xi  \exp (- \frac{\xi^2}{2 \gamma t} ) 
|\psi(\xi) \rangle \langle \psi(\xi)  | ,
\end{align}
which can be evaluated to be identical to (\ref{masterdyn}).

\begin{figure}
\scalebox{0.3}{\includegraphics{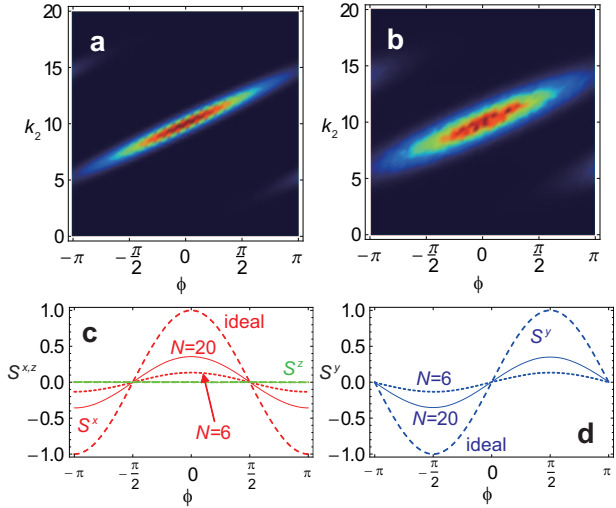}}
\caption{\label{fig5} 
Effects of dephasing on the teleportation protocol. (a) Measurement probability distribution (\ref{finalprob}) with no dephasing and (b) distribution (\ref{decofinalprob}) including dephasing.  Parameters used are $ N = 20 $, $ \gamma=1 $, $ k_1 = N $, $ k_3 = N/2 $, $ \theta = \pi/2 $. (c) (d) Bob's resulting state in the presence of dephasing for the particle numbers shown (solid and dotted lines), and the ideal case (dashed line). Parameters used are $ \gamma=1 $, $ \theta = \pi/2 $, $ k_1^{\mbox{\tiny cut}} =0 $. }
\end{figure}

Performing this procedure on the teleportation process Sec. \ref{secfull}, we obtain the following expression for Bob's state after the measurement Step 7
\begin{align}
\rho_{\mbox{\tiny Step7}} = & \frac{1}{2 \pi \gamma \sqrt{ T_2 T_3}}  \int d \xi \int d \xi' \exp (- \frac{\xi^2}{2 \gamma T_3} - \frac{\xi'^2}{2 \gamma T_2}) \nonumber \\
& \times | \cos (s_3 T_3 + \xi) e^{-i s_2 T_2 + \xi'}, \sin (s_3 T_3 + \xi) e^{i s_2 T_2 + \xi'} \rangle \rangle \nonumber \\
& \times \langle \langle \cos (s_3 T_3 + \xi) e^{-i s_2 T_2 + \xi'}, \sin (s_3 T_3 + \xi) e^{i s_2 T_2 + \xi'} | 
\label{dephasebobs}
\end{align}
where the measurement outcomes occurs with probability 
\begin{align}
 p_{\mbox{\tiny dec}} &(k_1,k_2,k_3)= \frac{1}{2 \pi \gamma \sqrt{ \tau_2 \tau_3}} \int d \xi \int d \xi'
\exp (- \frac{\xi^2}{2 \gamma \tau_3} - \frac{\xi'^2}{2 \gamma \tau_2}) \nonumber \\
&  \times  \frac{1}{4^N}C_N^{k_1}C_N^{k_2}C_N^{k_3} | A_{k_2 k_3} (\theta + 2\xi, \phi+ 2\xi')|^{2 k_1} \nonumber \\
& \times  | B_{k_2 k_3} (\theta + 2\xi, \phi+ 2\xi') |^{2N-2k_1}.
\label{decofinalprob}
\end{align}
The measurement probability distribution (\ref{decofinalprob}) without and with dephasing is plotted in Fig. \ref{fig5}(a) and \ref{fig5}(b) respectively.  We see that the dephasing broadens the probability distribution, a similar result to that found in the equatorial teleportation case \cite{pyrkov14}.  The effects of the dephasing is thus to loosen the correlations between Alice's original state and the measurement outcome, which in turn broadens the distributions of Bob's final state.  The dephasing also acts to dephase Bob's BEC qubit itself, as may be seen from (\ref{dephasebobs}).  These two effects contribute to a reduced fidelity of the teleportation protocol. 

Figs. \ref{fig5}(c)(d) show Bob's output state for two particle numbers $ N = 6,20 $ under dephasing.  We see that as with equatorial teleportation, the performance improves with $ N $.  As discussed in more detail in Ref. \cite{pyrkov14}, this is due to the nature of the entangling gates taking a physically shorter time $ \tau, T \sim 1/\sqrt{N} $,
thus resulting in less time for the dephasing to affect the state.  In the sense that the fidelity actually improves when $ N $ is increased shows the robustness of the protocol under decoherence.  This behavior would be the reverse if other types of entangled states were used, such as those involving Schrodinger cat states, which would behave catastrophically under dephasing. We thus find that the full Bloch sphere teleportation scheme as introduced in this paper has a similar behavior as equatorial teleportation under decoherence, with a favorable scaling towards large scale systems.

\section{Conclusions}

We have introduced a protocol that teleports a spin coherent state with an arbitrary position on the Bloch sphere to another spin coherent state via shared entanglement.  In contrast to the previously introduced equatorial teleportation scheme of Ref. \cite{pyrkov14} which required one auxiliary, or ancilla, BEC or spin ensemble, the scheme introduced here requires two ancilla BECs or ensembles. The ancilla spin coherent states are used to create two different quantum channels for transmitting the $\theta$ and $\phi$ spherical coordinates of Alice's original spin coherent state. As with standard qubit teleportation, classical information needs to be sent in order to ``correct'' the transmitted state.  We identified two types of classical bounds, one based on quantum state estimation, where Alice estimates her state the best she can, and sends this information to Bob.  Another bound, based on how much information is sent between Alice and Bob was also introduced.  The communication bound is broken unconditionally, but the quantum state estimation bound requires postselection to break the classical bound.  Under dephasing we find that the protocol scales favorably as the equatorial teleportation case, with the fidelity improving with increasing $ N $.  This is attributed to the use of entangled states of a time $ \tau,T = 1/\sqrt{2N} $ which are relatively robust against dephasing \cite{byrnes13}. 

The method explicitly works beyond the standard CV regime based on canonical position and momentum variables. Under the CV approximation, only spin coherent states in the vicinity of a polarized direction may be teleported.  The beyond-CV techniques, first introduced in Refs. \cite{byrnes12,pyrkov14} and generalized in this paper, allows for utilizing a larger portion of the total Hilbert space of the BECs and spin ensembles.  This allows for an alternative approach to performing quantum information processing, that is not within the standard qubit or CV formalism \cite{byrnes12}. Encoding quantum information on spin coherent states is attractive as contains a natural duplication of qubit information, making it robust against decohering processes such as loss. The addition of full Bloch sphere teleportation to the list of possible quantum algorithms possible using the framework \cite{byrnes12,pyrkov14,byrnes12b,pyrkov12,gross12} shows the potential of the approach as avenue towards realizing a quantum information processor.

\section*{Acknowledgments}
T. B. thanks Eugene Polzik, Mikhail Lukin, and Ujjwal Sen for discussions. This work is supported by the Transdisciplinary Research Integration Center, Non-MOU visiting grant NII, RFBR grant no. 14-07-31305, the Okawa Foundation, the Inamori Foundation, NTT Basic Laboratories, and JSPS KAKENHI Grant Number 26790061.


\end{document}